\documentclass[preprint2]{aastex}

\def\be{\begin{equation}}
\def\ee{\end{equation}}

\begin{document}

\title{The radiation belt of the Sun} 

\author{Andrei Gruzinov}

\affil{CCPP, Physics Department, New York University, 4 Washington Place, New York, NY 10003}

\begin{abstract}

For a given solar magnetic field, the near-Sun (phase-space) density of cosmic ray electrons and positrons of energy $\gtrsim 10$GeV can be calculated from first principles, without any assumptions about the cosmic ray diffusion. This is because the sunlight Compton drag must be more important than diffusion. If the solar magnetic field has an appreciable dipole component, the electron/positron density should have a belt-like dent, perhaps extending to several solar radii. The belt structure appears because the quasi-bound orbits are depopulated by the sunlight Compton drag. 

~~~

~~~

\end{abstract}

\section{ Introduction}

The sunlight photons can be Compton up-scattered by the cosmic-ray electrons and positrons. The resulting high-energy gamma-ray emission of the Sun has already been detected by EGRET ( Orlando \& Strong 2008) and Fermi (Abdo A. A. et al 2011) at photon energies $\gtrsim 100$MeV. The Sun might be detectable even at $\sim 100$GeV energies (Weiner et al 2013).

\begin{figure}
\plotone{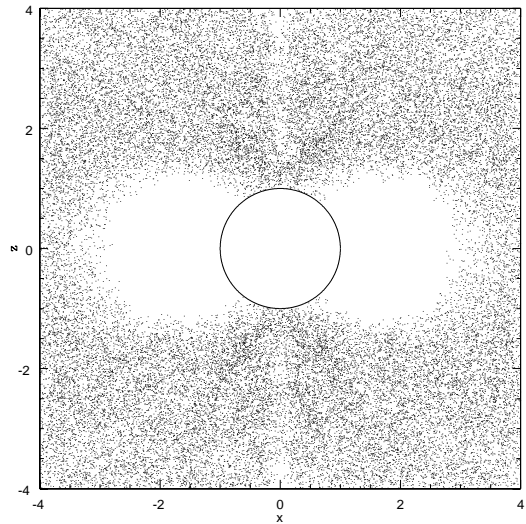}
\caption{Pure dipole field, 2G at the poles. Electron energy is 400GeV. Mean free path $\lambda =30{R_\odot^3\over r^2}$. Isotropic illumination from $10R_\odot$. The density of points is proportional to the electron number density. Smoothing with rms equal to ${R\odot\over 4}$ has been applied.}
\end{figure}

The inverse Compton gamma rays probe the near-Sun (phase-space) density of the cosmic-ray electrons and positrons. This density should have a non-trivial morphology. Here we note that the very existence of the Compton drag allows to calculate the near-Sun density of the high-energy electrons and positrons without any cosmic ray diffusion model. 

\section{ The Radiation Belt }

Given the solar magnetic field, the density of  $E\gtrsim 10$GeV electrons can be calculated just by a brute-force first-principle Monte-Carlo. One isotropically illuminates the Sun (from about ten solar radii is good enough) and calculates the numerically-exact trajectories. 

The result of the Monte-Carlo calculation is shown in Fig.1 for a pure-dipole magnetic field. To model the Compton drag, we removed the particles with a radius-dependent mean-free path $\lambda =30{R_\odot^3\over r^2}$ where $R_\odot$ is the radius of the Sun and $r$ is the spherical radius \footnote{ The particle removal is needed to explain why the bound orbits are empty. The Monte-Carlo generated density is actually insensitive to $\lambda$ for $\lambda \gtrsim {R_\odot^3\over r^2}$. The bound orbits are not really populated by the particles coming from afar, because of a nearly perfect conservation, at small radii $r$, of the adiabatic invariant $p_\perp^2/B$, where ${\bf p}_\perp$ is the momentum component perpendicular to the magnetic field ${\bf B}$.}.

To be clear, Fig.1 is a totally unrealistic toy model of the actual density around the Sun. In particular, for a pure dipole field, it does not make sense to discuss how the shape and size of the belt scale with the electron energy \footnote{The equatorial extent of the belt $r_e$ is determined by the breaking of the adiabatic invariant: $r_e\propto E^{-1/2}$, where $E$ is the electron energy. The belt boundary is parallel to the field line: $r\propto \sin ^2\theta$, where $\theta$ is the polar angle.}. The dipole approximation in never accurate, and must become totally misleading already at a few solar radii. 

But Fig.1 does correctly show that the quasi-bound orbits are practically empty, because the Compton drag depopulates these orbits faster than they are populated by the incoming electrons. Even if the solar field is not usefully approximated by the dipole, some magnetic field lines surely do emanate and terminate at the solar surface. In the vicinity of such field lines there will be some quasi-bound orbits, and these orbits are going to be empty.

\section{Conclusions}

\begin{itemize}

\item Given the (model) magnetic field in the vicinity of the Sun, one can calculate the near-Sun phase-space density of the high-energy electrons and positrons without any cosmic ray diffusion model -- from a first-principle Monte-Carlo. 

\item Generically, one expects a reduced electron density with non-trivial morphology -- the radiation belt of the Sun.

\item The radiation belt of the Sun might be detectable in the solar inverse Compton gamma rays.

\end{itemize}

\acknowledgements

I thank Neal Weiner and Itay Yavin for telling me about the solar high-energy gamma rays.

\end{document}